\documentclass{ws-procs9x6}
\newcommand{\bra}[1]{\langle {#1} |}
\newcommand{\ket}[1]{| {#1} \rangle}

\newcommand{\vecr}{{\mathbf r}}

\begin{document}

%%%%%%%%%%%%%%%%%%%%%%%%%%%%%%%%%%%%%%%%%%%%%%%%%%%%%%%%%%%%%% 
% title, author(s) and address(es) put here:                 %
%%%%%%%%%%%%%%%%%%%%%%%%%%%%%%%%%%%%%%%%%%%%%%%%%%%%%%%%%%%%%% 

\title{Response in the continuum for light deformed neutron-rich nuclei
\footnote{\uppercase{T}his work is supported by \uppercase{J}apan \uppercase{S}ociety for the \uppercase{P}romotion of
\uppercase{S}cience (1470146 and 14540369).}}

\author{Takashi Nakatsukasa\footnote{\uppercase{P}resent address:
\uppercase{I}nstitute of \uppercase{P}hysics,
   \uppercase{U}niversity of \uppercase{T}sukuba,
   \uppercase{T}sukuba 305-8571, \uppercase{J}apan.}}
\address{Department of Physics, Tohoku University, Sendai 980-8578, Japan}

\author{Kazuhiro Yabana}
\address{Institute of Physics, University of Tsukuba, Tsukuba 305-8571, Japan}

%%%%%%%%%%%%%%%%%%%%%%%%%%%%%%%%%%%%%%%%%%%%%%%%%%%%%%%%%%%%%%
% You may repeat \author \address as often as necessary      %
%%%%%%%%%%%%%%%%%%%%%%%%%%%%%%%%%%%%%%%%%%%%%%%%%%%%%%%%%%%%%%

\maketitle

\abstracts{
The time-dependent Hartree-Fock calculation with a full Skyrme
energy functional has been carried out on the three-dimensional
Cartesian lattice space to study $E1$
giant dipole resonances (GDR) in light nuclei.
The outgoing boundary condition for the continuum states
is treated by the absorbing complex potential.
The calculation for GDR in $^{16}$O suggests a significant influence
of the residual interaction
associated with time-odd densities in the Skyrme functional.
We also predict a large damping for superdeformed $^{14}$Be
at the neutron drip line.
}

%%%%%%%%%%%%%%%%%%%%%%%%%%%%%%%%%%%%%%%%%%%%%%%%%%%%%%%%%%%%%
% The main text of your paper                               %
%%%%%%%%%%%%%%%%%%%%%%%%%%%%%%%%%%%%%%%%%%%%%%%%%%%%%%%%%%%%%

\section{Time-dependent approach to nuclear response in the continuum}

The quantum-mechanical problems are usually solved in the
energy (time-independent) representation.
Namely, we either solve an energy eigenvalue problem for bound states or,
for scattering states,
we calculate a wave function 
with a proper boundary condition at a given energy.
However, if one wishes to calculate physical quantities in a wide
energy region,
the time-dependent approach is very useful because a single time
propagation provides information for a certain range of energy.
Another advantage may be an intuitive picture provided by the
time evolution of the wave function.

In Ref.~\refcite{NY01}, we have calculated
molecular photoabsorption cross sections in the electronic continuum
by using the time-dependent and time-independent approaches.
The results indicate the capability and efficiency of the time-dependent
method.
In the present paper, we will show nuclear response calculations using
the same technique,
the time-dependent Hartree-Fock (TDHF) method combined with the
absorbing boundary condition (ABC).

The TDHF state consists of $A$-occupied single-particle wave functions,
$\{ \phi_i(\vecr,t) \}_{i=1,\cdots,A}$,
each of which is complex and has two components (spinor).
The three-dimensional (3D) Cartesian coordinate, $(x,y,z)$,
is discretized in rectangular lattice
and derivatives are estimated with the nine-point formula.
The time evolution is determined by
\begin{equation}
\label{time_evolution}
\phi_i(\vecr,t) = 
\exp\left(-i\int_0^t dt' h_{\rm HF}[\rho(t')]\right)\phi_i(\vecr,0),
\end{equation}
where the HF Hamiltonian, $h_{\rm HF}[\rho(t)]$, depends on
$\phi_i(\vecr,t)$ and their derivatives.
The initial state, $\{ \phi_i(\vecr,0) \}_{i=1,\cdots,A}$,
is chosen as the HF ground state wave function
perturbed by an instantaneous external field,
$\phi_i(t=0)=e^{ik\hat{F}}\phi_i^0$.
Here, the coupling parameter, $k$, can be arbitrary
but should be small
to validate the Fourier analysis (linear response approximation).
The time variable is also discretized in a small step $\Delta t$ and
the exponential operator in Eq.~(\ref{time_evolution}) is
approximated by the fourth-order expansion,
\begin{equation}
\exp\left(-i\int_0^{\Delta t} dt' h_{\rm HF}[\rho(t')]\right)
\approx \sum_{n=0}^4 \frac{1}{n!} \left(h_{\rm HF}[\rho(\Delta t/2]\right)^n .
\end{equation}
The time evolution of the expectation value of the external field,
$\bra{\Psi(t)} \hat{F} \ket{\Psi(t)}$, is computed, then we utilize
the Fourier transform to obtain the strength function,
$S(E)\propto E\sum_n \delta(E-E_n) |\bra{n} \hat{F} \ket{0}|^2$.

Next, let us discuss the treatment of the continuum.
During the time evolution of the TDHF state, a part of the wave function
may escape from the nuclear binding and become an outgoing wave.
In the time-dependent approach, it is very difficult to impose the
outgoing boundary condition (OBC) explicitly, because the outgoing particles
have different energies.
Furthermore, the boundary condition for non-spherical systems
requires involved computation.\cite{NY01}
Thus, instead of exactly treating the OBC,
we use the absorbing boundary condition (ABC)
which can be practically identical to the OBC.
In the ABC method,
the Cartesian coordinate space ($r<R_0$) is divided into two parts:
an interacting region ($r<R_c$)
and a non-interacting region ($R_c<r<R_0$).
We introduce a complex absorbing potential, $-i\tilde{\eta}(r)$,
active only in the non-interacting region.
Then, we solve the TDHF equation of motion, Eq.~(\ref{time_evolution}),
with the box boundary condition (BBC)
at $r=R_0$, $\phi_i(\vecr,t)|_{r=R_0}=0$ for all occupied orbitals.
This means that the treatment of the continuum simply requires
an addition of the absorbing potential to the HF Hamiltonian.
\begin{equation}
h_{\rm HF}[\rho(t)]\Big|_{\mbox{{\sc obc} at }r=R_c} \doteqdot
 \left( h_{\rm HF}[\rho(t)]-i\tilde{\eta}(r)\right)
    \Big|_{\mbox{{\sc bbc} at }r=R_0}.
\end{equation}
We have demonstrated accuracy and practicality of
the ABC method both in the time-dependent\cite{NY01,NY02-P1,NY02-P2,YUN02-P}
and the time-independent approaches.\cite{UYN02,UYN02-P}

\section{Applications}
\subsection{Giant dipole resonance in $^{16}$O}

Giant dipole resonance (GDR) in $^{16}$O has been studied
extensively.\cite{HW01}
Since $^{16}$O is a doubly-closed spherical nucleus,
the continuum RPA\cite{SB75} is applicable.
A microscopic calculation with the Skyrme energy functional
was done\cite{LG76} and showed
a two-peak structure for GDR in $^{16}$O.
However, the calculation is not fully self-consistent because
the residual Coulomb, spin-orbit, and $\sigma\cdot\sigma$ part of
the interaction are neglected.
According to the best of our knowledge,
the continuum RPA calculation with the Skyrme functional has
never been carried out fully self-consistently.
Therefore, it is worth while to perform a fully self-consistent
TDHF+ABC calculation
and to investigate effects of the neglected part
of the residual interaction.

\begin{figure}[ht]
\centerline{\epsfxsize=0.85\textwidth\epsfbox{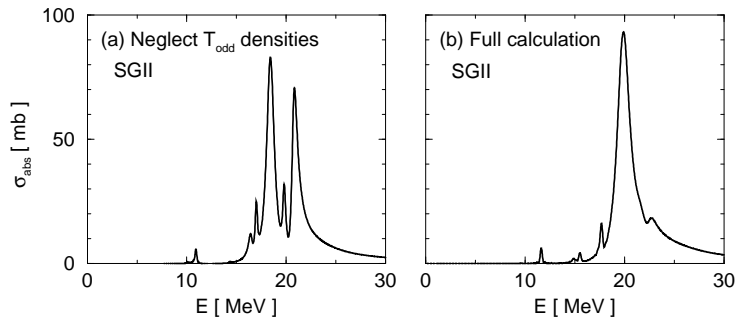}}
\caption{Photoabsorption cross section for $^{16}$O calculated with
different Skyrme energy functionals.
(a) Functional with SGII parameter set but neglecting
    all the time-odd densities.
(b) Full Skyrme functional with the SGII parameter set.
\label{GDR_O16}}
\end{figure}

We use the Skyrme energy functional of Ref.~\refcite{BFH87} with
the SGII parameter set.
The 3D sphere of radius $R_0=22$ fm is adopted as a model space.
The mesh spacing is $\Delta x\approx \Delta y \approx \Delta z\approx 1$ fm
inside the interacting region
but gradually increase up to $\Delta\approx 3$ fm in the non-interacting region.
After obtaining the HF ground state of $^{16}$O,
the isovector dipole field, $\hat{F}=z_n-z_p$,
is activated instantaneously at $t=0$.
The time evolution is computed up to $T=30\ \hbar/\mbox{MeV}$.
The result of the Fourier transform is shown in Fig.~\ref{GDR_O16}~(b).
Apparently, we observe only one peak at $E\approx 20$ MeV, except for
small peaks and a shoulder.
This is inconsistent with results of the continuum RPA.
The disagreement turns out to be due to the fact that
the continuum RPA neglects part of the residual interaction.
In the real-time calculation, we must use the same density
functional as the one to define the HF ground state.
Therefore, it is impossible to neglect the residual Coulomb and spin-orbit
interaction.
However, it is possible to neglect time-odd densities in the functional,
because this does not affect the ground state.
The calculation without the time-odd densities is shown in
Fig.~\ref{GDR_O16}~(a).
Now, the result becomes very similar to that of the continuum RPA
and there are two peaks in the continuum.
There are still some discrepancies related to the residual Coulomb
and spin-orbit force.
Our result indicates an important role of the time-odd density part of
the residual interaction.
Since the time-odd part of the Skyrme energy functional is not well
determined by the ground-state properties,
this might provide a useful constraint.
\subsection{Giant dipole resonance in $^{8,14}$Be}
\begin{figure}[ht]
\centerline{\epsfxsize=0.9\textwidth\epsfbox{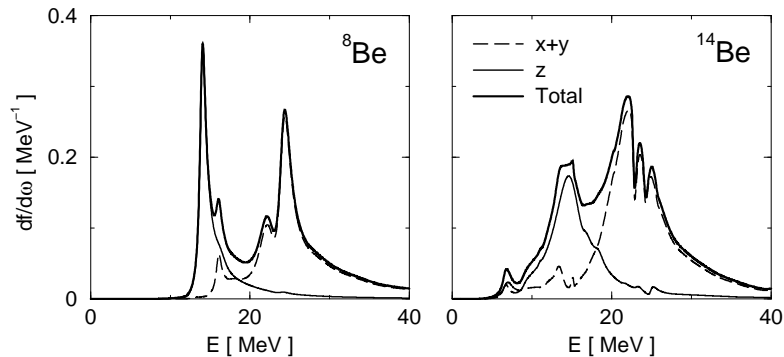}}
\caption{$E1$ Oscillator strength distribution for $^8$Be and $^{14}$Be
calculated with the full Skyrme energy functional with the SGII parameters.
The thin solid (dashed) line indicates the $E1$ oscillator strength
associated with the oscillation parallel (perpendicular)
to the symmetry axis.
\label{Be8_14}}
\end{figure}

Utilizing the same parameters and model space as $^{16}$O,
we discuss $E1$ resonances in light deformed nuclei,
$^8$Be and $^{14}$Be.
The quadrupole deformation of the HF ground states of these nuclei
are as large as $\beta\sim 1$ (prolate).
Calculated $E1$ oscillator strengths are shown
in Fig.~\ref{Be8_14}.
We can see a large deformation splitting of the GDR for $^8$Be.
The low-energy peak is sharper and located around 7 MeV of excitation energy.
This peak is associated with the isovector dipole oscillation along
the symmetry axis ($z$-axis).
The higher peak is rather broad and located around 22 MeV.
The difference in the broadening leads to a greater peak height
for the low-energy resonance,
although the integrated oscillator strength is twice larger for the
high-energy resonance.

For $^{14}$Be at the neutron drip line,
the average peak positions for low- and high-energy resonances are
similar to those in $^8$Be.
However, the excess neutrons significantly increase the peak width
of both resonances.
The large deformation splitting almost vanishes in the total
strength (thick solid line).
It is worth noting that the present calculation takes account of the
escape and the Landau damping width but does not include
the spreading width.
Some additional broadening might make the total strength look like
a single peak with a large damping.
%
%%%%%%%%%%%%%%%%%%%%%%%%%%%%%%%%%%%%%%%%%%%%%%%%%%%%%%%%%%%%%
%                                                           %
% You may repeat \section{SECTION N-th HEADING TYPE HERE}   %
%                                                           %
% Do start a subsection or sub-subsection, do this:-        %
%                                                           %
%   \subsection{SUBSECTION HEADING TYPE HERE}               %
%                                                           %
%   \subsubsection{SUBSUBSECTION HEADING TYPE HERE}         %
%                                                           %
% instead of the above                                      %
%                                                           %
%%%%%%%%%%%%%%%%%%%%%%%%%%%%%%%%%%%%%%%%%%%%%%%%%%%%%%%%%%%%%
%
\section{Conclusion}

We have carried out the Skyrme TDHF calculation with the ABC
in real time
for $E1$ resonances in the continuum.
The time-odd density components in the Skyrme energy functional
may influence considerably the isovector GDR strength distribution.
The GDR in Be isotopes have been studied and
the large deformation splitting and large broadening are predicted
for a drip-line nucleus, $^{14}$Be.

%%%%%%%%%%%%%%%%%%%%%%%%%%%%%%%%%%%%%%%%%%%%%%%%%%%%%%%%%%%%%
% Doing Acknowledgement                                     %
%%%%%%%%%%%%%%%%%%%%%%%%%%%%%%%%%%%%%%%%%%%%%%%%%%%%%%%%%%%%%
%\section*{Acknowledgments}

%%%%%%%%%%%%%%%%%%%%%%%%%%%%%%%%%%%%%%%%%%%%%%%%%%%%%%%%%%%%%
% Doing references:                                         %
%%%%%%%%%%%%%%%%%%%%%%%%%%%%%%%%%%%%%%%%%%%%%%%%%%%%%%%%%%%%%

%\bibliographystyle{unsrt}
%\bibliography{myself,nuclear_physics,chemical_physics}

%%%%%%%%%%%%%%%%%%%%%%%%%%%%%%%%%%%%%%%%%%%%%%%%%%%%%%%%%%%%%
%                                                           %
% Command to used is:-                                      %
%                                                           %
%  \bibitem{REFERENCE_LABEL} AUTHORS NAMES,                 %
%  {\it JOURNAL'S NAMES}{\bf VOLUME NUMBER}, PAGE (YEAR).   %
%                                                           %
%  See example below.                                       %
%                                                           %
%%%%%%%%%%%%%%%%%%%%%%%%%%%%%%%%%%%%%%%%%%%%%%%%%%%%%%%%%%%%%

\end{document}